\documentclass[11pt,preprint]{aastex}
\usepackage{chngpage}
\usepackage{graphicx}
\usepackage{natbib}

\shorttitle{Fitting to the multi-band afterglows of GRB 100814A}
\shortauthors{Yu et al. 2014}

\begin{document}

\title{Signature of a spin-up magnetar from multi-band afterglow rebrightening of GRB 100814A}

\author{Y. B. Yu\altaffilmark{1, 2}, Y. F. Huang\altaffilmark{1, 2}, X. F. Wu\altaffilmark{3, 4, 5},
M. Xu\altaffilmark{6}, J. J. Geng\altaffilmark{1, 2}}

\altaffiltext{1}{Department of Astronomy, Nanjing University, Nanjing 210093, China; hyf@nju.edu.cn}
\altaffiltext{2}{Key Laboratory of Modern Astronomy and Astrophysics (Nanjing University), Ministry of Education, China}
\altaffiltext{3}{Purple Mountain Observatory, Chinese Academy of Sciences, Nanjing 210008, China; xfwu@pmo.ac.cn}
\altaffiltext{4}{Chinese Center for Antarctic Astronomy, Chinese Academy of Sciences, Nanjing 210008, China}
\altaffiltext{5}{Joint Center for Particle Nuclear Physics and Cosmology of Purple Mountain Observatory-Nanjing University, Chinese Academy of Sciences, Nanjing 210008, China}
\altaffiltext{6}{Department of Physics, Jiangxi Science and Technology Normal University, Nanchang 330013, China}

\begin{abstract}
In recent years, more and more gamma-ray bursts with late rebrightenings in multi-band afterglows unveil the late-time activities of the central engines. GRB 100814A is a special one among the well-sampled events, with complex temporal and spectral evolution. The single power-law shallow decay index of the optical light curve observed by GROND between 640 s and 10 ks is $\alpha_{\rm opt} = 0.57 \pm 0.02$, which apparently conflicts with the simple external shock model expectation. Especially, there is a remarkable rebrightening in the optical to near infrared bands at late time, challenging the external shock model with synchrotron emission coming from the interaction of the blast wave with the surrounding interstellar medium. In this paper, we invoke a magnetar with spin evolution to explain the complex multi-band afterglow emission of GRB 100814A. The initial shallow decay phase in optical bands and the plateau in X-ray can be explained as due to energy injection from a spin-down magnetar. At late time, with the falling of materials from the fall-back disk onto the central object of the burster, angular momentum of the accreted materials is transferred to the magnetar, which leads to a spin-up process. As a result, the magnetic dipole radiation luminosity will increase, resulting in the significant rebrightening of the optical afterglow. It is shown that the observed multi-band afterglow emission can be well reproduced by the model.
\end{abstract}

\keywords{gamma rays: bursts -- ISM: jets and outflows -- individual: GRB 100814A}

\section{Introduction}
\label{sect:intro}

After the successful launch of the $Swift$ satellite, more and more cosmological Gamma Ray Bursts (GRBs) with complex behaviors, such as multiple X-ray flares or significant optical rebrightenings are detected (for a recent review, see Zhang 2007). For giant X-ray flares, GRB 121027A and GRB 111209A are the most interesting ones among the GRB samples.
The X-ray afterglow brightness of GRB 121027A increased by more than two orders of magnitude at about $10^{3}$ s after the trigger (Evans et al. 2012).
GRB 111209A is the longest-duration burst with a significant X-ray bump and a remarkable rebrightening in the optical band (Gendre et al. 2013). A lot of examples in GRBs with significant rebrightening in optical bands have also been observed, such as GRBs 970508 (Sokolov et al. 1999), 060206 (${\rm W\acute{o}zniak}$ et al. 2006), 081029 (Nardini et al. 2011), etc.
Considering the fact that X-ray flares share lots of common features with prompt emission, they are usually explained as due to internal shocks (Burrows et al. 2005; Fan \& Wei 2005; Zhang et al. 2006). However, the X-ray bumps of GRB 121027A and GRB 111209A are so special that they can not be explained by usual internal shock models. Wu et al. (2013) proposed a fall-back accretion model in the framework of the collapsar scenario (Woosley 1993; Paczynski 1998; MacFadyen \& Woosley 1999) to explain the sharp rebrightening of GRB 121027A observed at X-ray wavelength. Yu et al. (2013) applied the fall-back accretion model to GRB 111209A and successfully interpreted the unusual optical and X-ray afterglow light curves. For the remarkable rebrightenings in optical band, a simple external shock model with synchrotron emission coming from the forward shock fails to explain these observed complex behaviors. The energy injection model (e.g., Dai \& Lu 1998; Rees \& ${\rm M\acute{e}se\acute{a}ros}$ 1998; Huang et al. 2006; Dall'Osso et al. 2011; Yu \& Huang 2013), the two-component jet model (e.g., Huang et al. 2004; Liu et al. 2008), and the microphysics variation mechanism (e.g., Kong et al. 2010) are very popular in view of the fact they can explain the exceptional optical rebrightenings well.

GRB 100814A is another special event, with an early-time shallow decay phase and a late-time significant rebrightening in its optical afterglow light curve (Nardini et al. 2014). The power-law ($f_{\nu} \propto t^{-\alpha}$) temporal index of the early shallow decay is $\alpha = 0.57 \pm 0.02$, which is inconsistent with the external shock model expectation.
It is argued that the shallow decay phases come from the energy injection. Usually, the injection luminosity is assumed as $L(t) \propto t^{-q}$ (Nousek et al. 2006; Zhang et al. 2006; Yu \& Huang 2013), which may naturally come from the magnetic dipole radiation of a new-born millisecond magnetar (Dai \& Lu 1998; Zhang \& ${\rm M\acute{e}se\acute{a}ros}$ 2001; Dall'Osso et al. 2011). As a result, magnetars have been suggested as the central engines for some GRBs, including both long and short events (Zhang \& ${\rm M\acute{e}sz\acute{a}ros}$ 2001; Troja et al. 2007; Metzger et al. 2011; Bernardini et al. 2012; Rowlinson et al. 2013). In both Dai \& Lu's (1998) and Dall'Osso et al.'s (2011) work considering a strongly magnetized neutron star as the central engine of a GRB, the energy injection power is more realistically derived as $L(t) = L_{0}(1+t/T)^{-2}$, where $T$ is the spin-down timescale and $L_{0}$ is the initial luminosity. Especially, considering the exact form for the energy injection power of a spinning down magnetar due to magnetic dipole radiation, Dall'Osso et al. (2011) found that the luminosity of X-ray afterglow naturally has a shallow decay phase with a temporal power-law index of $\alpha \approx 0.5$. Recently, a nearly constant dipole radiation luminosity ($q \simeq 0$) during the spin-down timescale was favored by observations from GRBs, such as 050801 (de Pasquale et al. 2007), 060729 (Grupe et al. 2007), 080913 (Greiner et al. 2009). However, some observations of GRB afterglows with rebrightenings or bumps (i.e., $\alpha < 0$) require that the injection luminosity increases with time (i.e., $q < 0$). Additionally, there is a plateau phase in the X-ray band of GRB 100814A between $10^{3}$ s and $10^{5}$ s (Nardini et al. 2014), also indicating a continuous energy injection from the central engine during this prolonged period.

In this study, we suggest that the multi-band afterglow behavior of GRB 100814A can be well explained by considering the spin evolution of a central magnetar. The optical shallow decay phase and the X-ray plateau are due to energy injection from the magnetar in its early spin-down stage. The significant optical rebrightening observed at late time naturally comes from the spin-up process of the magnetar, which is caused by subsequent fall-back accretion.

Our paper is organized as follows. We summarize the observational facts of GRB 100814A in Section 2. The spin evolution of the magnetar during the fall-back accretion process, including the magnetic dipole radiation, is described in Section 3. In Section 4, we calculate the dynamics and radiation of the GRB afterglow external shock by considering energy injection from the central magnetar with spin evolution, and fit the unusual X-ray and optical afterglow light curves of GRB 100814A. We summarize our results and give a brief discussion in the final section. We assume a concordance cosmology of $H_{0}$ = 71 km ${\rm s^{-1}~Mpc^{-1}}$, $ \Omega_{M} $ = 0.27 and $\Omega_{\Lambda}$ = 0.73 throughout the paper.

\section{Observations}
\label{sect:obs}

GRB 100814A was detected by Burst Alert Telescope (BAT) onboard the $Swift$ observatory at 03:50:11 UT on October 14 2010 (Beardmore et al. 2010) and the duration measured with BAT is $T_{90} = 174.5 \pm 9.4$ s. The position of GRB 100814A was localized at $RA(J2000)=01^{h}29^{m}55^{s}, Dec(J2000)=-17^{\circ}59^{'}25.7^{''}$ (Krimm et al. 2010). The X-Ray Telescope (XRT) onboard the $Swift$ satellite began to observe GRB 100814A 87 s after the BAT trigger, when the gamma-ray emission was still detectable by BAT. GRB 100814A also triggered the Gamma-Ray Burst Monitor (GBM) onboard the $Fermi$ telescope with a duration of $T_{90} = 149 \pm 1$ s. The 1.024-sec peak photon flux measured in the GBM energy range (10 - 1000 keV) is $4.5 \pm 0.2 {\rm ~ph~s^{-1}~cm^{-2}}$. Given the fluence of $ f = (1.98 \pm 0.06) \times 10^{-5} {\rm ~erg~cm^{-2}}$ measured by GBM (von Kienlin 2010) and the redshift $ z = 1.44 $ reported by O'Meara et al. (2010), we can get the luminosity distance and the isotropic energy released in the rest frame as $D_{\rm L} = 10.5 \rm ~Gpc$ and $E_{\rm iso} = 1.04 \times 10^{53} \rm ~erg$ respectively.

\subsection{X-ray afterglow}

After the third peak observed by XRT at about 145 s, the X-ray afterglow light curve evolved into the so called steep decay phase, which is usually explained as the high latitude emission of a relativistic outflow at the true end of the prompt phase (Kumar \& Panaitescu 2000). The steep decay phase lasts until about 630 s, and is followed by a shallow decay phase from $\sim630$ s to about $2.0 \times 10^{4}$ s (see Fig. 1). During the shallow decay phase, the power law decay index is $\alpha_{\rm X,1} = 0.52 \pm 0.05$, which disagrees with the simple external shock scenario that the impulsive ejecta is expanding in an uniform medium or a stellar wind. As suggested by Nardini et al. (2014), a possible solution is to invoke a continuous energy injection into the ejecta. At about $10^{5}$ s, the X-ray afterglow light curve entered a steeper decay phase with a decay index of $\alpha_{\rm X,2} = 2.1 \pm 0.1$ until $\sim2 \times 10^{6}$ s. Interestingly, note that the X-ray emission remained constant after $2 \times 10^{6}$ s. It was interpreted as contribution from a nearby source (Nardini et al. 2014).

\subsection{Optical Afterglow}

UVOT onboard the $Swift$ satellite started to follow up GRB 100814A 80 s after the BAT trigger and a bright optical candidate was detected (Gronwall \& Saxton 2010). The Gamma-Ray burst Optical and Near-infrared Detector (GROND) mounted on the 2.2 m MPG/ESO telescope began observing GRB 100814A 150 s after the trigger and a bright optical source in all seven optical to NIR bands was detected. The index of the initial shallow decay phase from $\sim630$ s to $\sim2.0 \times 10^{4}$ s in all seven optical bands observed by GROND is $\alpha_{\rm opt,1} = 0.57 \pm 0.02$, which is consistent with the single power-law decay index measured with UVOT. The late-time optical afterglow showed unusual behavior, with a significant rebrightening from $\sim2.0 \times 10^{4}$ s in all seven bands (see Fig. 2). The optical flux measured by GROND increased by a factor of about 4 in about $8.0 \times 10^{4}$ s, interrupting the early time smooth temporal evolution. The remarkable late time rebrightening was also detected in all UVOT filters. Around the optical peak time of $10^{5}$ s, there is a hint of X-ray variability, but the amplitude of the rebrightening is much shallower compared with the optical. Considering the large error bars, a simultaneous rebrightening in X-ray is inconclusive. After the significant rebrightening, the optical afterglow light curve entered the quick decay phase with a decay index of $\alpha_{\rm opt,2} = 2.25 \pm 0.08$ (Nardini et al. 2014). This phase lasted until about $10^{6}$ s, where the contribution of the underlying host galaxy became dominant, which made the light curves flatten significantly.

\section{Fall-back Accretion and Magnetar Spin Evolution}
\label{sect:model}

Magnetars are a type of pulsars with strong dipole magnetic fields that exceed $4.4 \times 10^{13}$ G (Usov 1992; Duncan \& Thompson 1992). It is proposed that magnetars can be formed in the core-collapse of massive stars (Duncan \& Thompson 1992; Thompson \& Duncan 1993), or they could result from double white dwarf mergers (Usov 1992). It is also predicted that a magnetar can be formed before the remnant of a neutron star - neutron star merger collapses to a black hole (Rosswog et al. 2003; Price \& Rosswog 2006). Evidences for the existence of magnetars from neutron star - neutron star binary mergers are accumulating from observations (Norris et al. 1991; Rowlinson et al. 2010) and simulations (Giacomazzo \& Perna 2013; Kiuchi et al. 2014). Dall'Osso et al (2015) further investigated the gravitational wave emission from massive magnetars produced through magnetic field amplification during the binary neutron star mergers.

Recently, Dai \& Liu (2012) considered a newborn rapidly rotating magnetar surrounded by a hyperaccreting fall-back disk. They argued that the fall-back accretion process can make the magnetar spin up, which then leads to a strong energy injection due to enhanced dipole radiation. This mechanism can naturally account for the shallow decay phase, plateaus, and significant brightenings in GRB afterglows. In this paper, we assume a magnetar as the central engine of GRB 100814A to interpret the observed unusual optical and X-ray afterglow emission. We argue that the shallow decay phase in optical band and the plateau at X-ray wavelength is due to energy injection from a spin-down magnetar, which is initially losing its rotational energy through magnetic dipole radiation mechanism (Zhang \& ${\rm M\acute{e}sz\acute{a}ros}$ 2001; Dai 2004; Yu \& Dai 2007; Mao et al. 2010). At late time, when the magnetospheric radius, which is defined by the pressure balance between the fall-back material and the magnetic dipole, is smaller than the co-rotation radius, where the Keplerian angular velocity is equal to the rotation angular velocity of the central magnetar, materials will flow onto the surface of the magnetar. With the angular momentum of the accreted matter transferred to the magnetar, the latter will spin up and the magnetic dipole radiation luminosity increases, resulting in a significant rebrightening in the afterglow light curve.

Initially, the magnetar spins down through magnetic dipole radiation and the luminosity as a
function of time is
\begin{equation}
L_{\rm dip} = \frac{\mu^{2}{\Omega_{\rm s}}^{4}{\sin^{2}{\chi}}}{6c^{3}},
\end{equation}
where $\mu$ is the dipole magnetic moment and $\Omega_{\rm s}$ is the angular velocity of the magnetar. $\chi$ is the inclination angle of the rotation axis to the magnetic axis and we take a typical value
of $\sin^{2}{\chi} = 0.5$ in our calculations. With the magnetic dipole radiation, we can obtain the torque $\tau_{\rm dip}$ as
\begin{equation}
\tau_{\rm dip}= - \frac{\mu^{2}{\Omega_{\rm s}}^{3}\sin^{2}{\chi}}{6c^{3}}.
\end{equation}

Later on, the ejected materials whose kinetic energy is less than the potential energy will eventually fall back onto the central magnetar. In our calculations, the fall-back accretion start time, which is defined as when the fall-back accretion starts, is assumed to be $10^{4}$ s, which is derived from the beginning of the optical rebrightening of GRB 100814A. During the fall-back accretion process, the angular momentum of fall-back matter will be transferred to the central magnetar, which will lead the magnetar to spin up and increase the magnetic dipole radiation luminosity. The expression for the accretion torque $\tau_{\rm acc}$ is given by Dai \& Liu (2012) as
\begin{equation}
\tau_{\rm acc}=n(\varepsilon,\omega)\frac{\mu^{2}}{r_{\rm m}^{3}},
\end{equation}
where $n(\varepsilon,\omega)$ is the dimensionless torque parameter, and $r_{\rm m}$ is the magnetospheric radius defined as
\begin{equation}
r_{\rm m}=\left(\frac{\mu^{4}}{GM{\dot{M}}^{2}}\right)^{1/7},
\end{equation}
with $M$ and $\dot{M}$ being the magnetar mass and the mass rate of the fall-back accretion respectively.

Combining the contributions from spin-down and spin-up together, we can calculate the spin evolution of
the magnetar by
\begin{equation}
\frac{d(I\Omega_{s})}{dt}=\tau_{\rm dip}+\tau_{\rm acc},
\end{equation}
where $I$ is the moment of inertia. After assuming a constant moment of inertia, Dai \& Liu (2012) analytically obtained the relation between spin evolution and mass accretion rate. During the fall-back process, the mass accretion rate initially increases with time as $\dot{M} \propto t^{1/2}$ (MacFadyen et al. 2001; Zhang et al. 2008) and at late time the accretion rate decreases with time as $\dot{M} \propto t^{-5/3}$ (Chevalier 1989). So the spin evolution at early and late times are $\Omega_{s} \propto t^{23/28}$ and $\Omega_{s} \propto t^{-5/7}$ respectively, indicating that the magnetar spins up at early times and spins down at late times during the fall-back process.

In our study, the central engine of GRB 100814A is assumed to be a newly born magnetar with spin evolution. The early time optical shallow decay phase and X-ray plateau come from the initial spin-down process and the late time optical rebrightening results from the enhanced energy injection generated by spinning up of the magnetar. When calculating the dynamic process of the multi-band afterglow of GRB 100814A, we use the equations for beamed GRB outflows developed by Huang et al (1999, 2000) by considering continuous energy injection from the central magnetar with spin evolution. The bulk Lorentz factor ($\gamma$) of the shocked interstellar medium is described by the following differential equation (for details, see Huang et al 1999):
\begin{equation}
\frac{d\gamma}{dt}=\frac{-(\gamma^{2}-1)}{M_{ej}+\epsilon
m+2(1-\epsilon)\gamma m}\frac{dm}{dt},
\end{equation}
where $M_{ej}$ is the initial ejecta mass, $m$ is the swept-up interstellar medium mass, and $\epsilon$ is the radiative efficiency. When the energy injection due to magnetic dipole radiation from the central magnetar is taken into account, the above differential equation can be modified to be (Kong \& Huang 2010; Geng et al. 2013)
\begin{equation}
\frac{d\gamma}{dt}=\frac{1}{M_{ej}+\epsilon
m+2(1-\epsilon) \gamma
m}\times(\frac{1}{c^{2}}L_{\rm dip}-(\gamma^{2}-1) \frac{dm}{dt}).
\end{equation}
where $L_{\rm dip}$ can be calculated from Eq. (1). In our calculations, several effects, such as lateral expansion, electron synchrotron cooling and equal arrival time surfaces, have been incorporated. In the absence of electron synchrotron cooling, the comoving frame distribution of the shock-accelerated electrons is usually assumed to be a power-law function, $d{N_{\rm e}}^{'}/d\gamma_{\rm e} \propto {\gamma_{\rm e}}^{-p}$, where $p$ is the power-law index. After considering the electron synchrotron cooling effect, the distribution function will be changed to $d{N_{\rm e}}^{'}/d\gamma_{\rm e} \propto {\gamma_{\rm e}}^{-(p+1)}$ for electrons above a critical Lorentz factor $\gamma_{\rm c}$ (Sari et al. 1998). For detailed description of electron distribution, see Huang et al. (2000) and Huang \& Cheng (2003). We neglect the adiabatic pressure and energy losses due to adiabatic
expansion, which might also have a minor effect on the dynamical evolution of the blast wave (van Eerten et al. 2010; Pe'er 2012; Nava et al. 2013). For radiative process, the multi-band afterglow emission mainly come from synchrotron radiation of the shock-accelerated electrons due to their interaction with the magnetic field (Sari et al. 1998; Sari \& Piran 1999). We present our numerical results below.

\section{Numerical Results}
\label{sect:nume}

In our calculations, the central engine of GRB 100814A is assumed to be a rapidly rotating magnetar with a fall-back accretion disk. We argue that the optical and X-ray shallow decay phase is due to the re-energized process from a spin-down magnetar. When fitting the shallow decay phase, we take the same form of energy injection power as Dai \& Lu (1998). Considering the duration of the shallow decay phase in the X-ray afterglow of GRB 100814A, the spin-down timescale ($T$) is taken as $1.0 \times 10^{5}$ s. Other parameters are evaluated typically, such as the surface magnetic field strength $B_{0} = 1.0 \times 10^{15}$ G, initial rotation period $P_{0} = 2$ ms, which corresponds to an initial luminosity of $L_{0} = 1.5 \times 10^{47}~\rm erg~s^{-1}$. The simple external shock model apparently cannot explain the shallow decay phase observed in the multi-band afterglow. After considering the energy injection due to the spin-down process of the central magnetar, the shallow decay phase can be explained well. Note that we did not take the very early X-ray observational data into consideration in our fitting. In the very early X-ray afterglow light curve, there is a prompt peak, after which the X-ray afterglow light curve entered the sharp decay phase. The sharp decay can be explained as the tail emission after the prompt phase, while the peak may come from the contamination of the prompt emission.

After the shallow decay phase, a significant optical rebrightening appeared in the afterglow light curve at about $10^{4}$ s. We show that the optical rebrightening results from the increase of the magnetic dipole radiation when the materials from the accretion disk fall back toward the central magnetar, leading to a spin-up process. Following Wu et al. (2013), we numerically obtain the time evolution of the mass accretion rate. With the evolution of the mass accretion rate, we can get the spin evolution of the central magnetar for different sets of the model parameter values. To explain the strong late-time rebrightening observed in the optical band, the peak accretion rate is taken as $3.0 \times 10^{-6} M_{\odot}~\rm s^{-1}$, corresponding to a peak luminosity of $6.0 \times 10^{47}~\rm erg~s^{-1}$. The total mass accreted onto the magnetar during the whole fall back process is about $M_{\rm fb} \simeq 0.18 M_{\odot}$, which seems large when compared with the accreted mass in Wu et al. (2013). However, in this case the magnetar can still remain to be a neutron star and will not collapse to form a black hole. The observed optical and X-ray data of GRB 100814A and our best theoretical fit are illustrated in Figures 1 and 2 respectively. It is shown that the observed multi-band complex behavior of GRB 100814A can be well explained by invoking a newborn magnetar with spin evolution. Especially, the significant optical rebrightening can be satisfactorily reproduced. Around the optical peak time, the observational X-ray data show some hint of variability. Though the amplitude of the variation is much smaller than what is observed in optical bands, our model can also give a satisfactory fit. At the very late time, both X-ray and optical afterglow light curves are dominated by background emission. We interpreted the late optical emission as from the host galaxy, whose magnitude was estimated as $r^{'} \sim 30$ mag. We do not fit the very late X-ray afterglow light curve, which might be contaminated by a nearby source (Nardini et al. 2014). When fitting the optical afterglow light curves of GRB 100814A, we correct for the extinction of the host galaxy as summarized in Table 1.

\begin{figure}
   \begin{center}
   \plotone{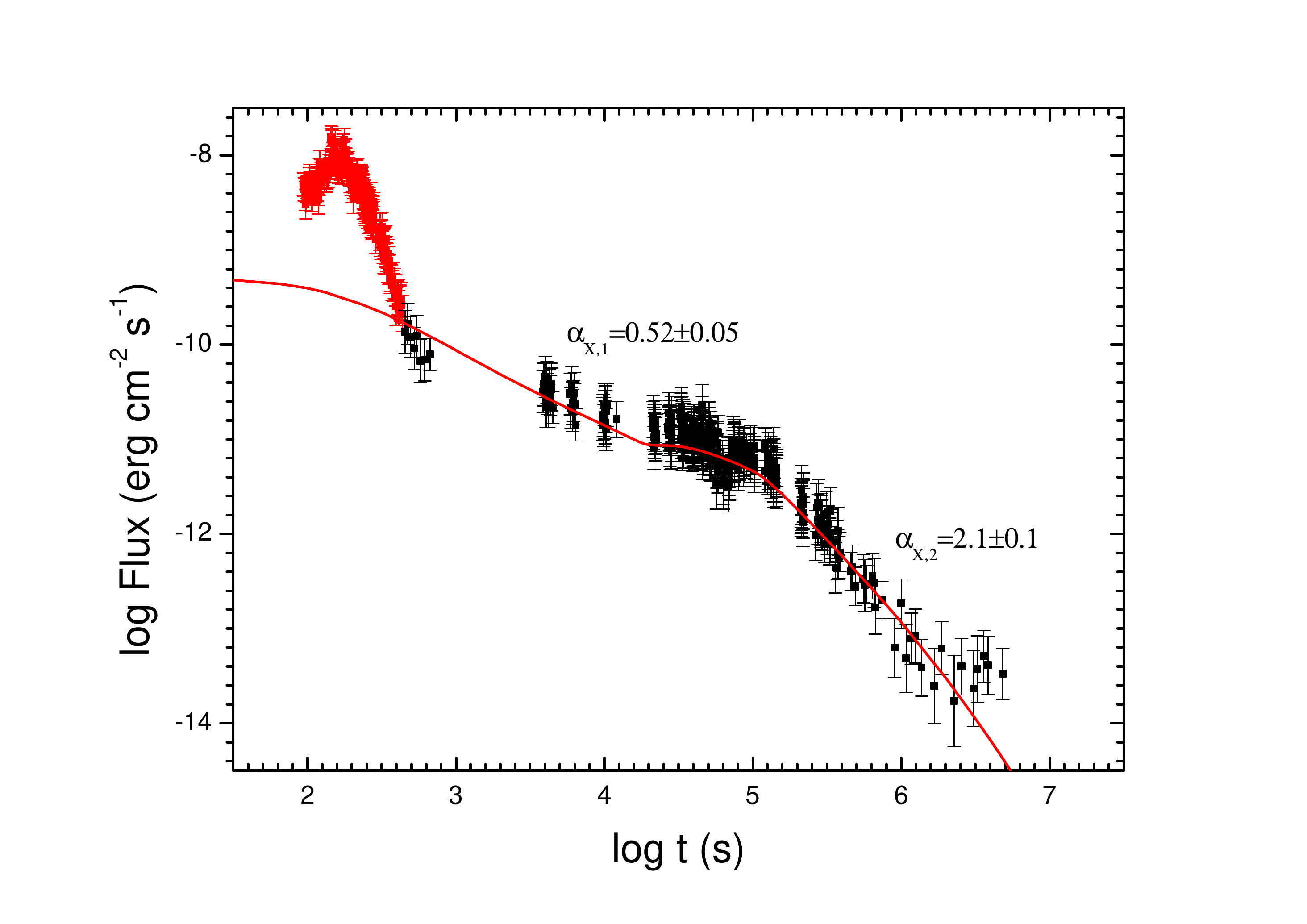}
   \caption{Theoretical fit to the X-ray afterglow light curve by considering a millisecond magnetar as the central engine of GRB 100814A. The red (Windowed Timing Mode) and black (Photon Counting Mode) points correspond to the observed XRT data (Nardini et al. 2014). The solid line represents our best fit by using the spin evolution model.}
   \label{Fig:plot1}
   \end{center}
\end{figure}

\begin{figure}
   \begin{center}
   \plotone{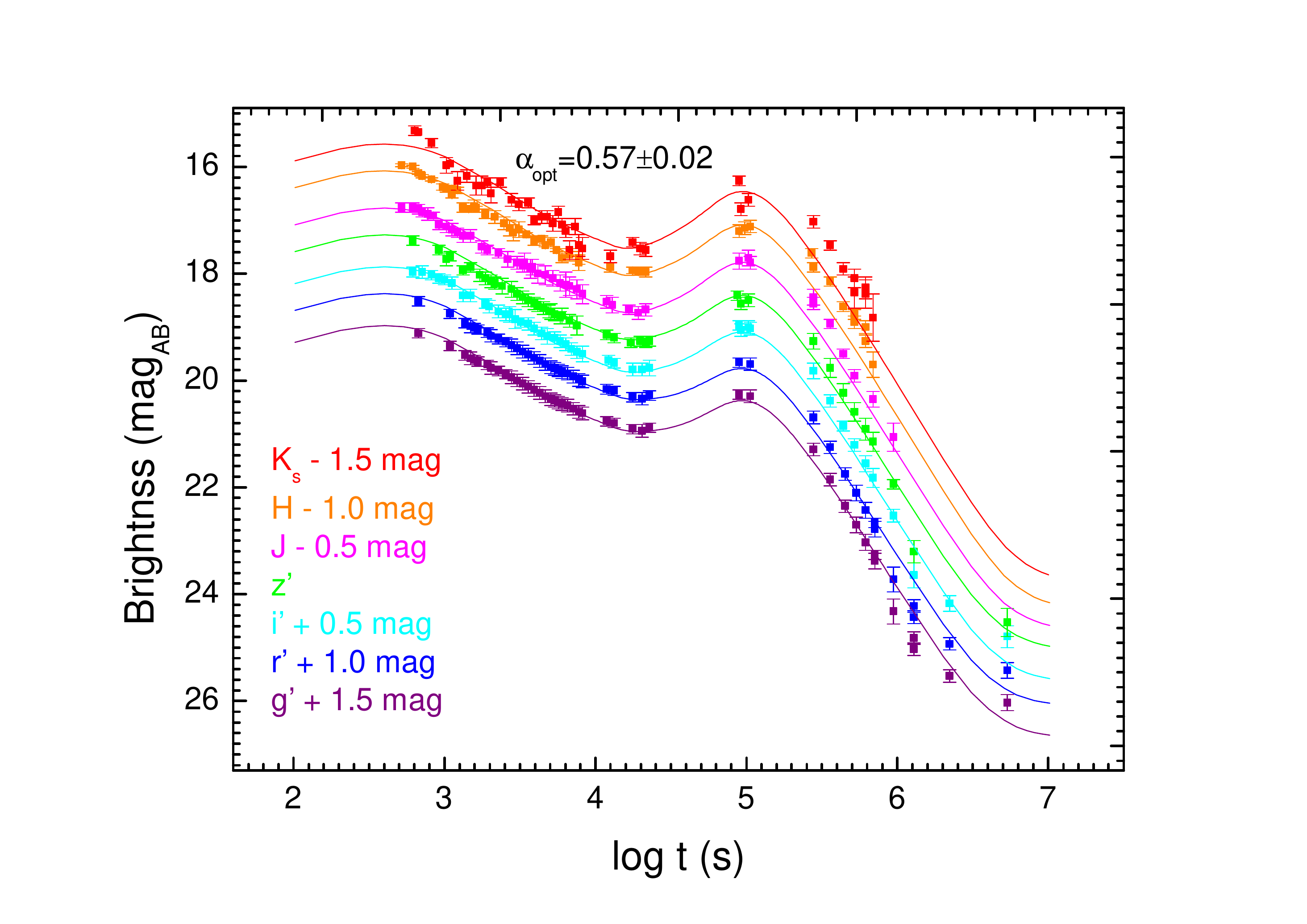}
   \caption{Numerical fit to the observed GROND seven-band optical afterglow light curves of GRB 100814A
   by using the same model parameters as in Figure 1. The points represent the observational
   data from Nardini et al. (2014). The solid lines correspond to our theoretical optical afterglow light curves with
   extinction corrected. A 0.5-magnitude shift between two adjacent light curves is applied in the plot for clarity.}
   \label{Fig:plot2}
   \end{center}
\end{figure}

\begin{deluxetable}{cccccccc}
\tabletypesize{\scriptsize}
\tablewidth{0pt}
\tablecaption{Dust extinction in seven bands adopted in our fit to GRB 100814A.\label{TABLE:Ext}}

\tablehead{%
        \colhead{} &
        \colhead{$g'$} &
        \colhead{$r'$} &
        \colhead{$i'$} &
        \colhead{$z'$} &
        \colhead{$J$} &
        \colhead{$H$} &
        \colhead{$K_{\rm s}$}}
\startdata
$A$ (mag) & 2.3 & 2.4 & 2.2 & 3.0 & 3.2 & 3.3 & 3.5 \\
\enddata
\end{deluxetable}

\section{Discussion and Conclusions}
\label{sect:disc}

A lot of different physical explanations have been proposed to interpret the shallow decay phase or the plateau phase of GRB afterglows. They include the energy injection mechanism, microphysics variation mechanism, two component jets, and so on (Zhang 2007). Among all these scenarios, the energy injection in the form of magnetic dipole radiation from a spin-down magnetar provides a natural explanation for the observed shallow decay phase or the plateau phase. Noting the late time optical rebrightening of GRB 100814A, we consider the effect of the fall back accretion on the spin evolution of the central magnetar. During the fall-back accretion process, angular momentum of the accreted materials is transferred to the magnetar and leads the latter to spin faster. It is the increase of magnetic dipole radiation that results in the significant rebrightening in the optical afterglow light curve of GRB 100814A.

As discussed in Yu \& Huang (2013), since the progenitors of long GRBs are associated with star forming regions/dusty molecular clouds, it is natural that there is an internal extinction in the host galaxies of cosmological GRBs, such as $A_{\rm V} \sim 2.5$ mag in GRB 970508 (Sokolov et al. 2001), $A_{\rm V} \sim 2.5$ mag in GRB 980703 (Kong et al. 2009), $A_{\rm V} > 2.5$ mag in GRB 050223 (Pellizza et al. 2006), $A_{\rm V} \sim 1.57$ mag in GRB 081029 (Yu \& Huang 2013) and $A_{\rm V} \sim 2.5$ mag in GRB 120804A (Berger et al. 2013). Recently, Covino et al. (2013) found that 13\% of the GRB afterglows are highly absorbed by computing rest frame extinction for a sample of GRB afterglows. For GRB 100814A, the adopted values of the dust extinction in GROND seven-bands are all in reasonable range.

Another distinguishing feature of GRB 100814A is the color evolution during the optical rebrightening. The rebrightening amplitude in the lower frequency bands is obviously higher than that in the higher frequency bands, which  makes our numerical results in J, H and $K_{\rm s}$ bands worse when compared with other bands. One possible solution to this problem is to invoke two different components during the rebrightening phase. A faster and narrower jet dominates the early afterglow light curve, while the late time light curve is dominated by a slower and wider jet. A color evolution is expected during the transition when the onset of the afterglow emission produced by the second jet occurs and the first jet begins to significantly diminish. The color evolution can also be explained as one of the characteristic frequencies, such as the cooling frequency, has crossed the observed bands under the assumption of non-evolving microphysical parameters in the external shock. Additionally, late prompt model (Ghisellini et al. 2007; Nardini et al. 2010), which assumes two components that are emitted in different regions: one component from the simple forward shock, and the other component related to a late-time activity of the central engine, is also a possible solution to the color evolution problem.

To explain the observed optical rebrightening of GRB 100814A, we assume the start time of the fall-back accretion to be $10^{4}$ s, which corresponds to a rest frame duration of $t_{\rm fb} \sim 4000$ s. Usually, the free-fall timescale is about $10^{3}$ s (Kumar et al. 2008). However, since the fall-back materials need to overcome the resistance of a neutrino-heated bubble (MacFadyen et al. 2001), the start time of the fall-back accretion may be a little larger than the free-fall timescale. Additionally, the ram pressure of the relativistic wind from the central magnetar may also affect the fall-back accretion (Dai \& Liu 2012). From the fall-back accretion start time in the rest frame, the maximal fall back radius of matter can be derived as $r_{\rm fb} \simeq 1.4 \times 10^{11} (M/1.4M_{\odot})^{1/3}(t_{\rm fb}/4000~\rm s)^{2/3}~\rm cm$.

During the fall-back accretion process, the accreted material will also liberate binding energy, producing an accretion luminosity, which will contribute to the overall emission. The ratio of accretion luminosity to the magnetic dipole radiation luminosity can be estimated as $k = \frac{GM\dot{M}/R}{\mu^{2}{\Omega_{\rm s}}^{4}{\sin^{2}{\chi}}/(6c^{3})} \sim 0.1 \left({\frac{B_{0}}{10^{15}~\rm G}}\right)^{-2}\left({\frac{R}{10^{6}~\rm cm}}\right)^{-7}$ for typical parameters, where $R$ is neutron star radius. So for GRB 100814A studied here, the accretion luminosity can be safely neglected in our calculations.

To summarize, it is shown that a newly born magnetar with spin evolution can reasonably explain both the optical and X-ray afterglow light curves of GRB 100814A. Especially the observed optical and X-ray shallow decay phase can be explained by energy injection in the initial spin-down process and the optical rebrightening can be reproduced quite well by assuming a fall-back accretion process, which leads the central magnetar to spin up and increases the magnetic dipole radiation luminosity.

\acknowledgments

\appendix
We acknowledge the referee for useful comments and suggestions. We thank Z.G. Dai and L.J. Wang for helpful discussions. This work was supported by the National Basic Research Program of China (973 Program, Grant No. 2014CB845800) and the National Natural Science Foundation of China (Grants Nos. 11473012, 11033002, 11322328 and 11203020). XFW acknowledges support by the One-Hundred-Talent Program, the Youth Innovation Promotion Association and the Strategic Priority Research Program ``The Emergence of Cosmological Structures'' (Grant No. XDB09000000) of Chinese Academy of Sciences and the Natural Science Foundation of Jiangsu Province (Grant No. BK2012890).

\end{document}